\documentclass[a4paper]{statsoc}

\usepackage[T1]{fontenc}
\usepackage[utf8]{inputenc}

\usepackage{fullpage}
\usepackage{natbib}

\usepackage{amscd,amsmath,latexsym,amssymb,amsfonts}
\usepackage{mathrsfs} 

\usepackage{bm}

\usepackage[a4paper,left=2cm,top=2cm,right=2cm,bottom=2cm,ignoreheadfoot]{geometry}

\title{Two discussions of the paper 
``Bayesian measures of model complexity and fit"
by D. Spiegelhalter et al., Read before The Royal Statistical 
Society at a meeting organized by the Research
Section on Wednesday, March 13th, 2002}
\author{El\'{\i}as Moreno$^{1}$, 
Francisco--Jos\'e V\'{a}zquez--Polo$^{2,3}$,
and Christian P.  Robert$^{4,5}$}
\address{
$^1$Universidad de Granada,
$^2$Universidad de Las Palmas de Gran Canaria,
$^3$TiDES Institute, Spain,
$^4$University of Warwick, UK,
and $^{5}$Université Paris-Dauphine, France}

\begin{document}
\maketitle 

\begin{abstract}
These are the written discussions of the paper 
``Bayesian measures of model complexity and fit"
by D. Spiegelhalter et al. (2002), following the
discussions given at the Annual Meeting of
the Royal Statistical Society in Newcastle-upon-Tyne
on September 3rd, 2013.
\end{abstract}

\section{Discussion by E.~Moreno and F.-J.~V\'{a}zquez--Polo}

This is an interesting paper, in which a new dimension correction to penalise
over-fit models is presented. It has given rise to considerable discussion;
here, we focus on the DIC model selection procedure defined in the paper.

\par Eleven years later, model selection for complex models remains an open
problem. The weak link of the Bayesian model selection approach is the
elicitation of the prior over models and over the model parameters to be used
in the procedure. Several priors have been proposed for interesting model
selection problems, such as variable selection in high dimensional regression,
clustering, change points and classification, but none of them satisfy all
reasonable requirements. Thus, we fully agree with the authors' claim in
justifying DIC that ``\textit{full elicitation of informative priors and
utilities is simply not feasible in most situations}''.  However, this does not
imply that in model selection we can avoid the use of priors in a coherent way
(Berger and Pericchi, 2001). 

\subsection{Does the DIC have a justification from a decision theory viewpoint?}

In model selection we have a sample $\mathbf{y}_{n}$ of size $n$, a discrete
class of $k$ competing sampling models $\mathfrak{M}$, the sampling density
of model $M_{i}$ is $f(\mathbf{y}_{n}|\theta _{i},M_{i})$, and a prior for
models and model parameters $\pi (\theta _{i},M_{i})=\pi (\theta _{i}|M_{i})$
$\pi (M_{i})$, where $\theta _{i}\in $ $\Theta _{i}\,$. The parameter spaces
are typically continuous.
\par
In model selection the quantity of interest is the model, and therefore the
decision space is $\mathfrak{D}=\big\{d_{j},$ $j=1,...,k \big\}$, where $d_{j}$ is
the decision to choose model $M_{j}$, and the states of nature is the
class of models $\mathfrak{M}$. Given a loss function $\mathfrak{L}(d_{i},M_{j}), \; \mathfrak{L}:\mathfrak{D\times M} \longrightarrow \mathbb{R}^{+}$, the optimal Bayesian decision is to choose the model $M^{\pi }$ such that 

\[
M^{\pi }=\arg \min_{i=1,...,k}\sum_{j=1}^{k}\mathfrak{L}(d_{i},M_{j})%
\pi (M_{j}|\mathbf{y}_{n}), 
\]%
\par\noindent
where 
$$\displaystyle 
\pi (M_{j}|\mathbf{y}_{n})=\frac{m_{j}(\mathbf{y}_{n})\pi (M_{j})}{%
\sum_{j=1}^{k}m_{j}(\mathbf{y}_{n})\pi (M_{j})}, 
$$
and the marginal $\displaystyle m_{i}(\mathbf{y}_{n})=\int_{\Theta _{i}}f(\mathbf{y}%
_{n}|\theta _{i},M_{i})\pi (\theta _{i}|M_{i})d\theta _{i},$ is
the likelihood of model $M_{i}, i=1,...,k$. This means that whatever loss
function $\mathfrak{L}(d_{i},M_{j})$ we use, the optimal decision depends on the posterior model
probabilities; that is, the decision formulation takes into account the
uncertainty of the model. However, the DIC does not depend on $\pi(M_j|\mathbf{y}_{n}), \; j=1,...,k.$

\subsection{Does the DIC correspond to a Bayesian procedure?}

The Bayesian procedures automatically penalise model complexity without
any adjustment (Dawid, 2002), and this is a good reason to require
a model selection procedure to be Bayesian. Another reason is that
the competing models can be averaged, with the weights being the model
posterior probabilities. On the other hand, for Schwarz's Bayesian information criterion (BIC), to compare model $M_{i}$ with $M_{j}$, 
\[
-2\log BIC_{ij}(\mathbf{y}_{n})=-2\log \frac{f(\mathbf{y}_{n}|\hat{%
\theta}_{i}(\mathbf{y}_{n}),M_{i})}{f(\mathbf{y}_{n}|\hat{\theta}_{j}(%
\mathbf{y}_{n}),M_{j})}+(d_{i}-d_{j})\log n, 
\]%
where $d_{i},d_{j}$ are the dimensions of $\Theta _{i}$ and $\Theta _{j}$,
there is a Bayes factor $B_{ij}$ such that $|-2\log~B_{ij} -  2 \log BIC_{ij}|=O_{P}(n^{-1/2})$ (Kass and Wasserman, 1995), and thus the $BIC$ asymptotically corresponds
to a Bayes factor, we do not see that a similar
correspondence can be established with the
\[
DIC_{ij}(\mathbf{y}_{n})=-2\log \frac{f(\mathbf{y}_{n}|\bar{\theta}%
_{i}(\mathbf{y}_{n}),M_{i})}{f(\mathbf{y}_{n}|\bar{\theta}_{j}(\mathbf{y}%
_{n}),M_{j})}+ \mbox{Correction}_{ij} 
\]%
where $\bar{\theta}_{i}(\mathbf{y}_{n})=E_{\theta _{i}|\mathbf{y}_{n}}\theta
_{i}$, and%
$$\displaystyle
\mbox{Correction}_{ij} = 4 \; \Big\{E_{\theta _{i}|\mathbf{y}_{n}}\log f(%
\mathbf{y}_{n}|\theta _{i},M_{i})-E_{\theta _{j}|\mathbf{y}_{n}}\log 
f(\mathbf{y}_{n}|\theta _{j},M_{i}) \Big\} 
+ \; 4\log \frac{f(\mathbf{y}_{n}|\tilde{\theta}_{i}(\mathbf{y}%
_{n}),M_{i})}{f(\mathbf{y}_{n}|\tilde{\theta}_{j}(\mathbf{y}_{n}),M_{j})}.
$$

We note that under mild conditions ${\Large |}\hat{\theta}(\mathbf{y}_{n})-%
\bar{\theta}(\mathbf{y}_{n}){\Large |}=O_{P}(n^{-1})$, and hence the main
difference between BIC and DIC comes from the correction term. As a result of this term, the DIC does not correspond to a Bayesian procedure.

\subsection{Asymptotic.}

The DIC is not a consistent model selection procedure and although
it is a negative property of the procedure, this does not seem to worry
the authors, who argue that ``\textit{we neither believe in a true model
nor would expect the list of models being considered}''. This 
implies that the probability of a model has no meaning, as no model
space is considered. However, the point is that if we applied the DIC to
a case in which the class of models were known, we would
have consistency.

On the other hand, some statisticians, for instance Fraser (2011), have
suggested that the sampling properties of the Bayesian
methods should be studied. In this respect, Wasserman (2011) asserts that ``we must
be vigilant and pay careful attention to the sampling properties of
procedures". We agree with both these views. Moreover,
consistency is a very useful sampling property that allows us to compare
the behaviour of alternative Bayesian model selection procedures for complex models.

Consistency in a model selection procedure for a given class of 
models $\mathfrak{M}$ means that when sampling from a model in $\mathfrak{M},$
the posterior probability of this model tends to one as the sample size tends to infinity.
Bayesian procedures for model selection are typically consistent
when the dimension of the models is small compared with the sample
size (David, 1992; Casella et al., 2009). Furthermore, when the model
from which we are sampling is not in the class $\mathfrak{M}$, the Bayesian procedure asymptotically
chooses a model in $\mathfrak{M}$ that is as close as possible  to the true one, in
the Kullback--Leibler distance. 

On the other hand, consistent Bayesian procedures for low dimensional
models are not necessarily consistent for high dimensional models. For example: (a) Schwarz's approximation to the Bayes factor BIC  is not necessarily consistent in high dimensional settings (Berger, 2003; Moreno et al., 2010). $\,$ (b) When the number of models increases with the sample size, as occurs in clustering, change point or classification problems, consistency
of the Bayesian model selection procedure depends not
only on the prior over the model parameters but also on the prior
over the models. In fact, default priors commonly used for discrete
spaces may give an inconsistent Bayesian model selection procedure,
as occurs in clustering when using the uniform prior over
the models (Casella et al., 2012). $\,$ (c) In variable selection in regression when the number of regressors $p$
increases with the sample size, i.e., $p=O(n^{b}), 0\leq b\leq 1$, some priors
that are commonly used over the model parameters and over the model space make the
Bayesian procedures inconsistent. For instance, the $g-$priors (Zellner, 1986)
with $g=n$ produce an inconsistent Bayesian procedure. The mixture of $g-$priors with respect
to the InverseGamma$(g|1/2, n/2),$ or the intrinsic priors (Moreno et al., 1998) over the model parameters
when combined with the independent Bernoulli prior on the model
space (George and McCulloch, 1997; Raftery et al., 1997)
may also provide an inconsistent Bayesian procedure.

\bigskip\par
These results show that consistency can be a very useful property
for the difficult task of selecting priors for model selection in
complex models. 

\section{Discussion by C.P.~Robert}

The main issue with DIC undoubtedly is the question of its worth for (or
within) Bayesian decision analysis (since I doubt there exist many proponents
of DIC outside the Bayesian community). The appeal of DIC is, I presume, to
deliver a {\em single} summary per model for all models under comparison and to
allow therefore for a complete ranking of those models. I however object at the
worth of simplicity for simplicity’s sake: models are complex (albeit less than
reality) and their usages are complex as well. To consider that model A is to
be preferred upon model B just because $DIC(A)=1228 < DIC(B)=1237$ is a mimicry
of the complex mechanisms at play behind model choice, especially given the
wealth of information provided by a Bayesian framework. (Non-Bayesian paradigms
may be more familiar with procedures based on a single estimator value.) And to
abstain from accounting for the significance of the difference between $DIC(A)$
and $DIC(B)$ clearly makes matters worse.

This is not even discussing the stylised setting where one model is considered
as ``true" and where procedures are compared by their ability to recover the
``truth". David Spiegelhalter repeatedly mentioned during his talk that he was
not interested in this. This stance inevitably brings another objection,
though, namely that models--as tools instead of approximations to reality--can
only be compared against their predictive abilities, which DIC seems unable to
capture. Once again, what is needed in this approach to model comparison is a
multi-factor and all-encompassing criterion that evaluates the predictive
models in terms of their recovery of some features of the phenomenon under
study. Or of the process being conducted.  (Even stooping down to a
one-dimensional loss function that is supposed to summarise the purpose of the
model comparison does not produce anything close to the DIC function, unless
one agrees to massive approximations.)

Obviously, considering that asymptotic consistency is of no importance
whatsoever (as repeated by David Spiegelhalter in his presentation) easily
avoids some embarrassing questions, except the (still embarrassing) one about
the true purpose of statistical models and procedures. How can those be
compared if no model is true and if accumulating data from a given model is not
meaningful? How can simulation be conducted in such a barren landscape? I find
this minimalist attitude the more difficult to accept that models are truly
used as if they were or could be true, at several stages in the process. It
also prevents the study of the criterion under model misspecification, which
would clearly be of interest.

Another point worth discussing, already exposed in Celeux et al.
(\citeyear{celeux:forbes:robert:titterington:2006})), is that there is no unique driving principle
for constructing DICs. In that paper inspired from the discussion by De Iorio
and Robert (\citeyear{deiorio:robert:2002}), we examined eight different and
natural versions of DIC for mixture models, resulting in highly diverging
values for DIC and the effective dimension of the parameter, I believe that
such a lack of focus is bound to reappear in any multi-modal setting and fear
that the answer about (eight) different focus on what matters in the model is
too cursory and lacks direction for the hapless practitioner.

My final and critical remark about DIC is that the criterion shares very much
the same perspective as Murray Aitkin’s integrated likelihood, as already
stressed in \cite{robert:titterington:2002}. Both Aitkin
(\citeyear{aitkin:1991}, \citeyear{aitkin:2010}) and Spiegelhalter et al.
(\citeyear{spiegbestcarl}) consider a posterior distribution on the
likelihood function, taken as a function of the parameter but omitting the
delicate fact that it also depends on the observable and hence does not exist a
priori. See Gelman et al. (\citeyear{gelman:robert:rousseau:2013})) for a
detailed review of Aitkin’s (2010) book, since most of the criticisms therein
equally apply to DIC, and I will not reproduce them here, except for pointing
out that DIC escapes the Bayesian framework (and thus requires even more its
own justifications).

\end{document}